\documentclass[12pt,preprint]{aastex}
\usepackage{epsfig}

\begin{document}
\title{Identifying silicate-absorbed ULIRGs at z$\sim$1--2 in the
  Bootes Field using Spitzer/IRS}

\author{M. M. Kasliwal\altaffilmark{1,2},
V. Charmandaris\altaffilmark{3,1,4}, D. Weedman\altaffilmark{1},
J.R. Houck\altaffilmark{1}, E. Le Floc'h\altaffilmark{5,4},
S.J.U. Higdon\altaffilmark{1},L. Armus\altaffilmark{6}, H. I. Teplitz\altaffilmark{6}}

\altaffiltext{1}{Astronomy Department, Cornell University, Ithaca, NY 14853, USA}
\altaffiltext{2}{Astronomy Department, California Institute of Technology, 105-24, Pasadena, CA 91125, USA}
\altaffiltext{3}{Department of Physics, University of Crete, GR-71003, Heraklion, Greece}
\altaffiltext{4}{Chercheur Associ\'e, Observatoire de Paris, F-75014, Paris, France}
\altaffiltext{5}{Steward Observatory, University of Arizona, 933 North Cherry Avenue, Tucson, AZ 85721, USA}
\altaffiltext{6}{Spitzer Science Center, California Institute
of Technology, 220-6, Pasadena, CA 91125, USA}

\email{mansi@astro.caltech.edu,vassilis@physics.uoc.gr,dweedman@isc.astro.cornell.edu,jrh13@cornell.edu,elefloch@as.arizona.edu,sjuh@astro.cornell.edu,lee@ipac.caltech.edu,hit@ipac.caltech.edu}

\begin{abstract}

  Using the 16$\mu$m peakup imager on the Infrared Spectrograph
  (IRS\footnote{The IRS was a collaborative venture between Cornell
    University and Ball Aerospace Corporation funded by NASA through
    the Jet Propulsion Laboratory and the Ames Research Center.}) on
  Spitzer, we present a serendipitous survey of 0.0392 deg$^{2}$ within
  the area of the NOAO Deep Wide Field Survey in Bootes. Combining our 
  results with the available Multiband Imaging Photometer for Spitzer (MIPS)
  24$\mu$m survey of this area, we produce a catalog of 150 16$\mu$m
  sources brighter than 0.18\,mJy (3$\sigma$) for which we derive
  measures or limits on the 16/24$\mu$m colors. Such colors
  are especially useful in determining redshifts for sources whose mid
  infrared spectra contain strong emission or absorption features that
  characterize these colors as a function of redshift. We find that
  the 9.7$\mu$m silicate absorption feature in Ultraluminous Infrared
  Galaxies (ULIRGs) results in sources brighter at 16$\mu$m than at
  24$\mu$m at z $\sim$ 1--1.8 by at least 20$\%$. With a threshold flux
  ratio of 1.2, restricting our analysis to $>5\sigma$ detections at
  16$\mu$m, and using a $3\sigma$ limit on 24$\mu$m non-detections,
  the number of silicate-absorbed ULIRG candidates is 36. This
  defines a strong upper limit of $\sim$920 sources deg$^{-2}$,
  on the population of silicate-absorbed ULIRGs at  z $\sim$ 1--1.8.  
  This source count is about half of the total number of sources 
  predicted at z $\sim$ 1--2 by various phenomenological models. 
  We note that the high 16/24$\mu$m colors measured cannot be reproduced 
  by any of the mid-IR spectral energy distributions assumed by these 
  models, which points to the strong limitations currently affecting our
  phenomenological and theoretical understanding of infrared galaxy
  evolution.

\end{abstract}



\section{Introduction}

It is widely accepted that in order to fully understand the observed
increase in star formation activity at high redshifts
\citep[e.g.][]{Madau98}, a more comprehensive understanding of the
ultraluminous infrared population (ULIRGs) will play a key
role. ULIRGs, galaxies with infrared luminosity L$_{IR} > 10^{12}$
L$_{\sun}$, are rare in the local universe and comprise only $3\%$ of
the IRAS Bright Galaxy Survey \citep{Soifer87}. Yet, at high redshifts
of z $> 2$, ULIRGs may account for the bulk of all star-formation
activity and dominate the far-infrared background
\citep[e.g.][]{Blain02}.  The interstellar dust formed in starburst
galaxies absorbs the optical and UV emission and re-radiates in the
mid and far infrared. These galaxies enshrouded in dust are extremely
difficult to directly observe in the optical and near infrared regime.
A number of theoretical groups,
\citet[e.g.][]{Lagache04,Chary04,Pearson05,Gruppioni05} have developed
galaxy evolution models that constrain the evolution of the infrared
luminosity function with redshift. These semi-empirical models predict
the comoving luminosity density distribution and mid-infrared source
counts as a function of redshift.

The complexities of the ULIRG spectra, the possibility that most could
contain contributions from an active galactic nucleus (AGN) and a
massive starburst \citep{Genzel98}, challenges the interpretations of
the mid-IR surveys especially prior to Spitzer's advent
\citep{Elbaz02,Fadda02}. The superb sensitivity of the IRS on Spitzer
\citep{Houck04} showed the diversity of the mid-IR spectra of ULIRGs
in the local universe \citep{Armus04,Spoon04}. Moreover, the imaging
capability of IRS at 16 and 22 $\mu$m to levels below $\sim$0.1\,mJy
in addition to the broadband filters of the Spitzer cameras have
allowed the use of mid-IR colors as tracers of specific spectral
features \citep[see][]{Charmandaris04b}.

One such mid-IR continuum feature is the 9.7$\mu$m silicate absorption
band. Since it is not prominent in normal galaxies, quasars or unobscured
starbursts, it can be used as an indicator of high columns of cold
dust obscuring the nuclear emission from dust rich IR luminous
systems. The presence of this feature has been clearly seen in the
local universe from the ground \citep[e.g.][]{Dudley99} as well as in
space with the Infrared Space Observatory (ISO)
\citep{Genzel98,Laurent00} and Spitzer \citep{Armus04, Spoon04}. It
has also been clearly detected in sources at higher redshifts such as
at z$\sim$1 \citep{Higdon04} and at z$\sim$2 \citep{Houck05}.

Mid-IR color anomalies due to this feature can be used as an
approximate redshift indicator of a high redshift IR luminous source,
if there is a large line of sight extinction to the nucleus. This is
important because spectroscopic redshifts are not readily available
for these distant and optically faint ULIRGs. Furthermore, at
z$\sim$1--2, determining redshifts is challenging due to the so-called
``redshift desert'' as strong UV/optical emission lines are not
accessible from the ground.  The variation of mid-IR colors in Spitzer
data due to the presence of the 9.7$\mu$m absorption feature was proposed
by \citet{Charmandaris04b} as a potential redshift indicator for SCUBA
sources. Subsequently \citet{Takagi05} presented a detailed analysis
on the effects of the 9.7$\mu$m silicate absorption feature on mid-IR
colors measured by Spitzer and ASTRO-F broadband filters as a function
of redshift. They predict a population of galaxies which they call
``Si-break'' galaxies. These are galaxies at z$\sim$1.5, which due to
strong 9.7$\mu$m absorption are not detected (or are extremely faint)
by the 24$\mu$m filter of the MIPS(\citep{Rieke04}), even though they
are more prominent at other mid-/far-IR wavelengths. Given the
sensitivity of the Spitzer instruments, as well as the expected SEDs
and redshift of these sources, these galaxies would have to be dust
enshrouded galaxies with L$_{IR}$ $>$ 10$^{12}$L$_{\sun}$ i.e. ULIRGs.

In this paper, we compare the 24$\mu$m observations of the $\sim$9
deg$^2$ NOAO Deep Wide-Field Survey \citep[NDWFS;][]{Jannuzi99} in
Bootes using Spitzer/MIPS, to the 16 $\mu$m peak-up imaging of
Spitzer/IRS obtained in parallel during deep spectroscopic
observations in the same area.  Our goal is to identify
silicate-absorbed ULIRGs in the Bootes field and compare it to
theoretical predictions of total number of sources at z $\sim$ 1--2.
This is the largest area to-date for which deep sub-mJy level imaging
at both 24$\mu$m and 16$\mu$m is available.

We present our observations in $\S$ 2, results in $\S$ 3, and discuss
the implications of our findings in $\S$ 4.

\section{Observations and Data Reduction}

We observed 57 positions in the Bootes field during two different
periods using the IRS on Spitzer \citep{Houck04}.  Note that the
peak-up images utilized for the present analysis were obtained in
parallel as a ``bonus'' during deep IRS staring spectroscopic
observations ($\sim$ 7--35 $\mu$m) of select sources in the Bootes
field \citep{Houck05}. When the IRS spectrum of a science target
between 7--15$\mu$m was obtained using the Short Low module, images of
two different parts of the sky one with the blue peak-up camera at
16$\mu$m (13.3--18.7 $\mu$m) and another with the red peak-up camera
at 22$\mu$m (18.5--26 $\mu$m) were acquired in parallel.

The total field of view of a blue peak-up image in both nod positions
is 50$\times$55 pixels. Since the pixel size of the IRS short-low (SL) 
module is $\sim1.8\arcsec\times1.8\arcsec$, the field of view is 2.475
arcmin$^2$.  The total area observed is 0.0392 deg$^2$ which is
$\sim$230 times smaller than the total area of the Bootes survey imaged at
24$\mu$m with MIPS.  The observations were obtained between August 27
and September 2, 2004 as well as between November 11 and 17, 2004,
with exposure times of $\sim$240sec, resulting in a median 1$\sigma$ depth of
$\sim$0.06mJy at 16$\mu$m. 

The IRS 16$\mu$m images were processed using the standard IRS pipeline
(version 11.0) at the Spitzer Science Center (see chapter 7 of Spitzer
Observing
Manual\footnote{http://ssc.spitzer.caltech.edu/documents/som/}). The
2D images were converted from slopes after linearization correction,
subtraction of darks, and cosmic ray removal. The resulting images
were divided by the photometric flat, and a world coordinate system
was inserted into them using the reconstructed pointing of the
telescope. The astrometric accuracy of our images is better than
$\sim$1$\arcsec$ and the FWHM of the point spread function (PSF) is
$\sim$3.5$\arcsec$ at 16$\mu$m. The peak-up images of each of the two
nod position were median averaged and the final images of the two nods
were subtracted from each other.This removed the background emission
to first order and facilitated the source identification, which was
performed by eye. 1$\sigma$ in sources detected in both nod positions 
was lower by a factor of $\sim \sqrt{2}$.

To calculate the exact location of the source, a Gaussian fit was used
to obtain its centroid.  Aperture photometry was performed in the
location of the centroid and the flux was measured within an aperture
radius of 3 pixels. A median sky was subtracted from an annulus of
inner and outer radii of 8 pixels and 17 pixels respectively. Final
conversion from counts (e$^{-}$/s) to mJy was done by dividing by 729,
a factor determined during the flux calibration of the peak-up arrays
after observing a variety of stars for which high quality spectral
templates were available. We estimate that our photometry is accurate
to a 6\% level for sources $>3\sigma$.

\section{Results}

The 16$\mu$m sources in the Bootes field are cataloged in Table 1.  We
detect a total of 150 sources at 16$\mu$m over the 0.0392 deg$^2$
region in the Bootes field with a flux greater than 3$\sigma$ of
$\sim$0.18\,mJy. Of these 150 sources, 137 have optical counterparts
available from the NOAO survey (within 2\arcsec) and 80 have 24$\mu$m
counterparts (within 2.5\arcsec). 

\begin{deluxetable}{cccccrccc}
\tabletypesize{\footnotesize}
\footnotesize
\setlength{\tabcolsep}{0.1in}
\tablecaption{Catalog of the 16$\mu$m sources in the Bootes NOAO Wide Deep Field}
\tablewidth{0pc}
\tablehead{
\colhead{ID}  & \colhead{Spitzer Name\tablenotemark{a}}  & \colhead{f$_{16 \mu m}$}  & \colhead{$\sigma_{16 \mu m}$} & \colhead{f$_{24 \mu m}$\tablenotemark{b}} &  \colhead{$\frac{\rm f_{16 \mu m}}{\rm f_{24 \mu m}}$}  & \colhead{$B_W$}  &  \colhead{$R$}  &  \colhead{$I$}  \\
  &   & \colhead{(mJy)}  & \colhead{(mJy)} & \colhead{(mJy)}  &  & \colhead{(mag)}  &  \colhead{(mag)}  &  \colhead{(mag)}  }
\startdata
  1    &   SST16 J143413.49+332217.4    &      0.275    &      0.057    &    \nodata    &   $>$ 1.52    &   \nodata    &    24.72     &    23.65    \\
  2    &   SST16 J143408.45+332218.1    &      0.331    &      0.048    &     0.346     &       0.95    &    22.18     &    20.85     &    20.32    \\
  3    &   SST16 J143411.11+332212.2    &      0.237    &      0.048    &     0.207     &       1.14    &    24.82     &    23.74     &    23.00    \\
\enddata

\tablenotetext{a}{SST16 source name derives from discovery with the IRS PU 16$\mu$m images; coordinates listed are in J2000; 16$\mu$m positions with typical 3\,
$\sigma$ uncertainty of $\pm$ 1.2\arcsec; sources with an optical counterpart
will also appear in NDWFS catalogs with prefix NDWFS and the optical source position; 
optical magnitudes are Mag-Auto from NDWFS Data Release Three, available at 
http://www.noao.edu/noao/noaodeep/); sources with a MIPS 24 $\mu$m counterpart 
will also appear in MIPS catalog.}
\tablenotetext{b}{Ratio of the 16$\mu$m and 24$\mu$m flux densities. Upper limit
s for 24$\mu$m sources are assumed to be 0.18 mJy.}

\tablecomments{Table 1 is published in its entirety in the electronic
edition. A portion is shown here for guidance regarding its form and
content.}

\end{deluxetable}

To better quantify the presence of silicate-absorbed ULIRGs at
z$\sim$1.5 in the Bootes field, as proposed by \citet{Takagi05}, we
plot the distribution of sources as a function of the ratio of
16$\mu$m to 24$\mu$m flux in Figure~1.  For a 0.3\,mJy 
source, the uncertainty in the f$_{16 \mu m}$/f$_{24 \mu m}$ ratio using error 
propagation is $\sim$ 0.3. Therefore, we restrict our analysis to the
67 16$\mu$m sources brighter than 0.3\,mJy (5$\sigma$). For sources
that do not have 24$\mu$m counterparts in the Bootes catalog, we
manually inspected the location corresponding to the 16$\mu$m sources
and in several cases we were able to identify faint sources below the
formal 0.18\,mJy (3$\sigma$) limit of the catalog. For non-detections
we compute a lower limit to the ratio by assuming a 3$\sigma$ limit of
0.18\,mJy to the 24$\mu$m flux. We find 30 matched and 18 unmatched
sources that are brighter at 16$\mu$m relative to 24$\mu$m.  Of these
sources, 18 matched and 18 unmatched sources have f$_{16 \mu m}$/f$_{24 \mu m}$ ratio greater than 1.2.

\section{Discussion and Conclusion}

To examine in more detail the variation of the 16$\mu$m to 24$\mu$m
color, following the approach of \citet{Charmandaris04b}, we plot in
Figure~2 the ratio of 16$\mu$m flux to the 24$\mu$m flux based on
Spitzer IRS spectra of ULIRGs with strong or moderate silicate
absorption. We also include in the plot an average starburst mid-IR
SED as well as an AGN and a quasar. These galaxies were selected after
careful examination of over 120 mid-IR spectra in the 5--38$\mu$m
range obtained as part of the IRS guaranteed time extragalactic
program. This sample provides the best coverage of parameter space for
the integrated mid-IR SEDs of galaxies available to date.  We find
that the 9.7$\mu$m silicate absorption feature clearly causes a peak
in the 16$\mu$m to 24$\mu$m flux ratio, when the emitting source is at
z$\sim$1.5 because it is in the center of the 24$\mu$m band.  From
Figure~2, when the depth of the silicate absorption increases, the
peak of the ratio and the width of the redshift range over
which the emitting galaxy can be located also increase. Other mid-IR
emission features such as the strong 7.7$\mu$m feature, attributed to
the C--C stretch mode of polycyclic aromatic hydrocarbons (PAH) seen
in many starburst galaxies \citep[e.g.][]{Forster03,Brandl05}, are too
weak to push this ratio above 1.2.

Conservatively, we set the threshold for ULIRGs with strong
9.7$\mu$m silicate absorption at f$_{\rm 16\mu m}$/f$_{\rm 24\mu
  m}>$ 1.2. Therefore, restricting our analysis to 16$\mu$m detections
$>5\sigma$ (f$_{\rm 16\mu m}>$ 0.3\,mJy), we identify 18 sources with
24$\mu$m counterparts and 18 with no counterparts as silicate-absorbed
ULIRGs in 0.0392 deg$^2$, i.e.  $\sim$ 920 sources deg$^{-2}$. 
Based on our available mid-IR SEDs and Figure~2, this would
set an upper limit to the possible redshift range of these galaxies of
z $\sim$ 1--1.8.  Our identification does not include ULIRGs with
warmer SEDs such as Mrk1014 or Mrk231 \citep{Armus04,Weedman05}. The
above number places a {\it strong upper limit} on the population
of silicate-absorbed ULIRGs at the redshift epoch of z $\sim$ 1--1.8.

\citet{Takagi05} predict $\sim$ 900 silicate-break galaxies deg$^{-2}$
for their bright end model and $\sim$ 1500 deg$^{-2}$ for their burst
model. Their prediction is based on f$_{\rm 16\mu m}$/f$_{\rm 22\mu
  m}>$0.8 and includes models with deep silicate absorption. We chose
a higher threshold of 1.2 to minimize contamination. As is evident
from Figure 2, a threshold of 0.8 would select starburst galaxies as
well as prototypical AGNs like Mrk 231. From Figure 1, we find
$\sim$1450 deg$^{-2}$ above the \citet{Takagi05} threshold of 0.8. This
puts a strict upper limit on the population of silicate-break
galaxies.

How do our results compare with other theoretical predictions? Given
that we are interested in sources with f$_{16 \rm \mu m}$/f$_{24 \rm
  \mu m}$$>$1.2 and that most of the model predictions in the literature
are for the 24$\mu$m surveys we focus on predictions to the {\it total
  number of sources} with f$_{24 \rm \mu m}>$0.2\,mJy in the redshift
range of z $\sim$ 1--2.  According to \citet{Lagache04} these are $\sim$1732
deg$^{-2}$ while \citet{Gruppioni05} predict $\sim$1828
deg$^{-2}$. The ``burst'' and ``bright end'' models of
\citet{Pearson05} result in 1644 to 1550 deg$^{-2}$ and
\citet{Chary04} predict $\sim$1663 deg$^{-2}$.  Interestingly, the
ensemble of the mid-IR SEDs which are being used by the theoretical
models to fit the number counts and produce the above mentioned
predictions do not include SEDs which have an extreme 9.7$\mu$m band
such as IRASF00183-7111.

Yet, our study finds that heavily absorbed sources such as
IRASF00183-7111 are sufficient to account {\it for more than half of
  all} the sources at z $\sim$ 1--2 predicted by these various
models. Examining the 10 ULIRGs in the Bright Galaxy Sample,
\citet{Armus05} find that half of them would exhibit a ratio f$_{\rm
  16\mu m}$/f$_{\rm 24\mu m}>1.2$. In this context our upper limit of
920 ULIRGs with strong silicate absorption at z$\sim$1--1.8 is
consistent with what one would expect based on the ULIRGs in the local
universe. The theoretical models currently available cannot make a
direct prediction on the number of these types of galaxies at high
redshift since they are based on observations of only a handful of
mid-IR SEDs and the wealth of Spitzer/IRS spectra which clearly
demostrate the diversity of the mid-IR features, are only now becoming
available in the literature. Interestingly, \citet{Lagache04} show
that small variations in the shape of the PAH emission features of the
earlier work of \citet{Lagache03} were necessary to explain the
increase in the 24$\mu$m number counts detected by the Spitzer deep
surveys. Similarly, one would expect that taking into account the new
mid-IR SEDs of ULIRGs may have a significant influence on the
theoretical predictions of the type of infrared luminous galaxies
contributing to the observed number counts.

It is clear that there are some caveats on the above mentioned
approach as a method for identifying sources with strong silicate
absorption at z $\sim$ 1--2. Inspection of Table 1, suggests that all our
sources, with the exception of source \#33 and \#41, have an I-mag
greater than 19 and an R-mag greater than 20. If any of these sources
were stars, their I and R magnitudes are inconsistent with what we
find. For instance, if we consider a main sequence star whose 16$\mu$m
flux is 0.3\,mJy, its V-band magnitude varies from 11.2 (type B0,1V)
to 17.5 (type M,late V) \citep[see ][]{Wainscoat92}. This corresponds
to a range in R band magnitude of 11.3-15.7. Clearly, the R-mag of all
our sources is 4--5 mags fainter. Thus it is highly unlikely that we
are observing faint main sequence stars. Red giants, embedded
protostars or asteroids are also improbable contaminants because of
the high Galactic and high Ecliptic latitute of the Bootes field
(b$\sim$67$\deg$ and l$\sim$45$\deg$ respectively) as well as the
multiple epochs of the MIPS24 catalogue.

Since there are no direct spectroscopic observations available for at
least a fraction of the galaxies of our sample, we cannot
cross-calibrate the method using broadband colors and the exact
redshift of the source.  However, \citet{Teplitz05} performed a
similar study around Hubble Deep Field North, using an area $\sim$4
times smaller than the present study but with considerably deeper
imaging at 24 $\mu$m and with spectroscopic redshifts for most of
their sources. Their results reveal that 10 of their 149 sources have
known redshift and f$_{\rm 16\mu m}$/f$_{\rm 24\mu m}>1.2$, half of
which are at z$>$1. 20 sources with ratio greater than 1.2 have
unknown redshifts. Assuming that the mid-IR and optical redshift cross
identifications are accurate, these could be considered as an
approximate estimate for the uncertainties. Even though we have no
indication that we are incomplete in the mid-IR SED sampling, there
is always a possibility that SEDs which are not taken into account
here, are affecting our estimates. For example, it is conceivable that
a population of quiescent dwarf galaxies with low optical luminosity
consistent with the NOAO R-mag $\sim$20 located at z$\sim$0.2--0.5
and mid-IR spectra similar to those of the spiral nucleus of M51
\citep[see ][]{Teplitz05} could produce such a ratio. It is also
possible that some complications may arise in our sample if some of
our sources are unresolved interacting galaxies having components with
significantly different mid-IR spectra \citep[see
][]{Charmandaris04a}.

Irrespective of possible contamination contributing to an overestimate
of our observed counts, our upper limit illustrates the importance of
silicate-absorbed ULIRGs as a substantial fraction of the ULIRG
population at z $\sim$ 1--2. The faint brightness level of these
galaxies (f$_{24\mu m}<$0.5mJy) makes the direct detection of the
mid-IR spectral shape for a substantial sample of them rather
challenging even for Spitzer/IRS. As a result, infrared broadband
imaging and accurate SED fitting techniques using local analogues may
be the only method to provide constraints to the ambiguities in the
theoretical predictions to the redshift distribution of the sources
contributing to the IR number counter. The recent addition of
efficient 16$\mu$m imaging with Spitzer as well as the upcoming
ASTRO-F mission will clearly play a critical role towards this goal.

\acknowledgments 

This work is based on observations made with the Spitzer Space
Telescope, which is operated by the Jet Propulsion Laboratory,
California Institute of Technology, under NASA contract 1407. Support
for this work was provided by NASA through Contract Number 1257184
issued by JPL/Caltech. The authors would like to thank R. Chary and
B.T. Soifer (SSC/Caltech) for valuable discussions, as well as the
anonymous referee whose suggestions improved the manuscript.


\clearpage

\begin{figure}
\epsscale{0.5}
\centerline{\includegraphics[angle=90]{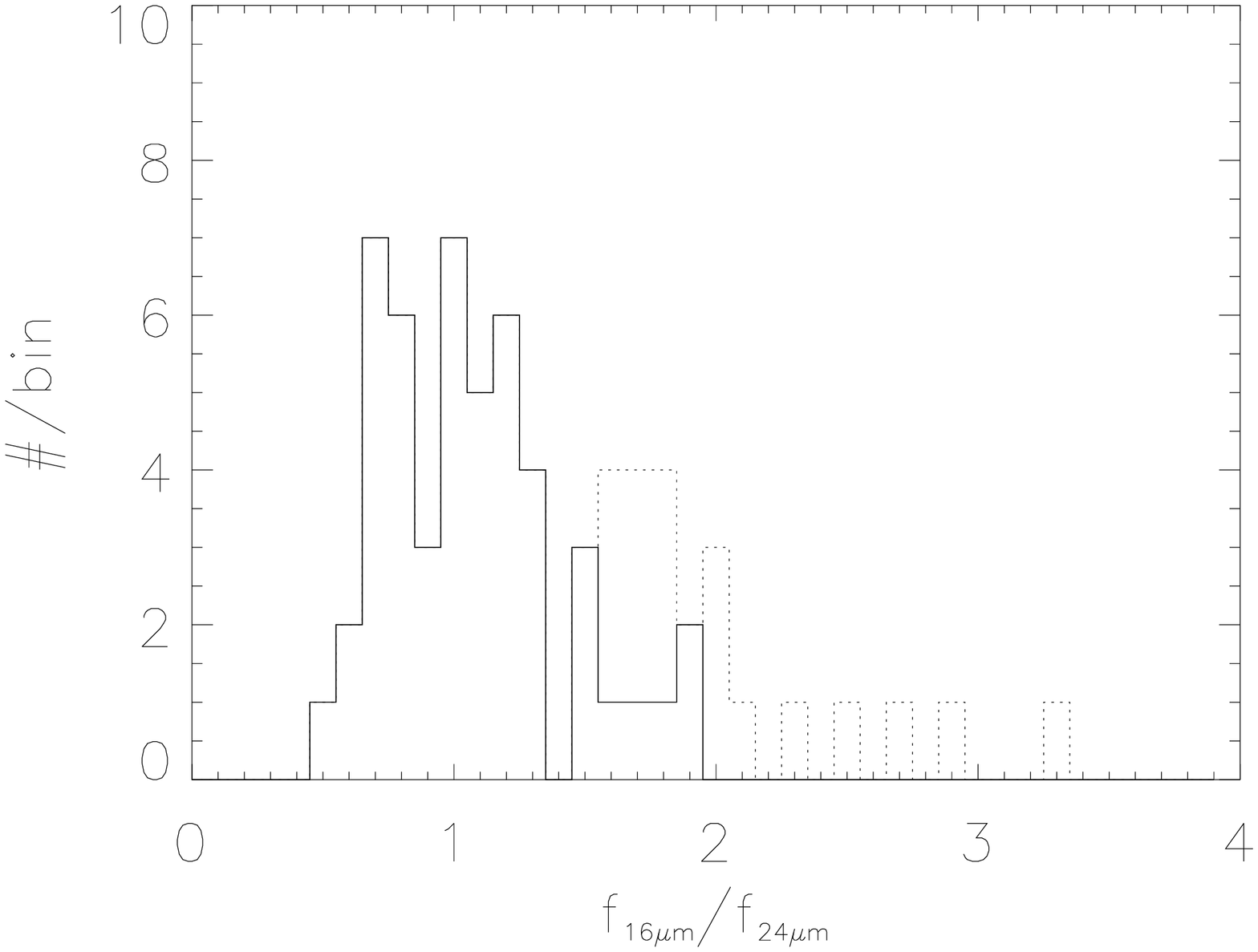}}
\caption{A histogram of 16$\mu$m over 24$\mu$m flux density ratio for
  all sources with f$_{\rm 16\mu m} > $0.5 mJy. The solid line
  represents sources found in both catalogs. Dashed lines represents
  the increase in number of sources if we assume a lower limit on
  ratios for 16$\mu$m sources with no 24$\mu$m counterpart.  These
  ratios are computed based on the 0.18\,mJy limit for MIPS24 $\mu$m
  imaging \citep[see ][]{Houck05}. We define as potential
  silicate-absorbed ULIRGs all sources with f$_{\rm 16\mu m}$/f$_{\rm
    24\mu m}>$1.2. As seen from Fig. 2, depending on the strength of
  the silicate band, these systems can be located at redshifts z
  $\sim$ 1--1.8.\label{histogram}}
\end{figure}

\clearpage
\begin{figure}
\epsscale{0.6}
 \centerline{\includegraphics[angle=90]{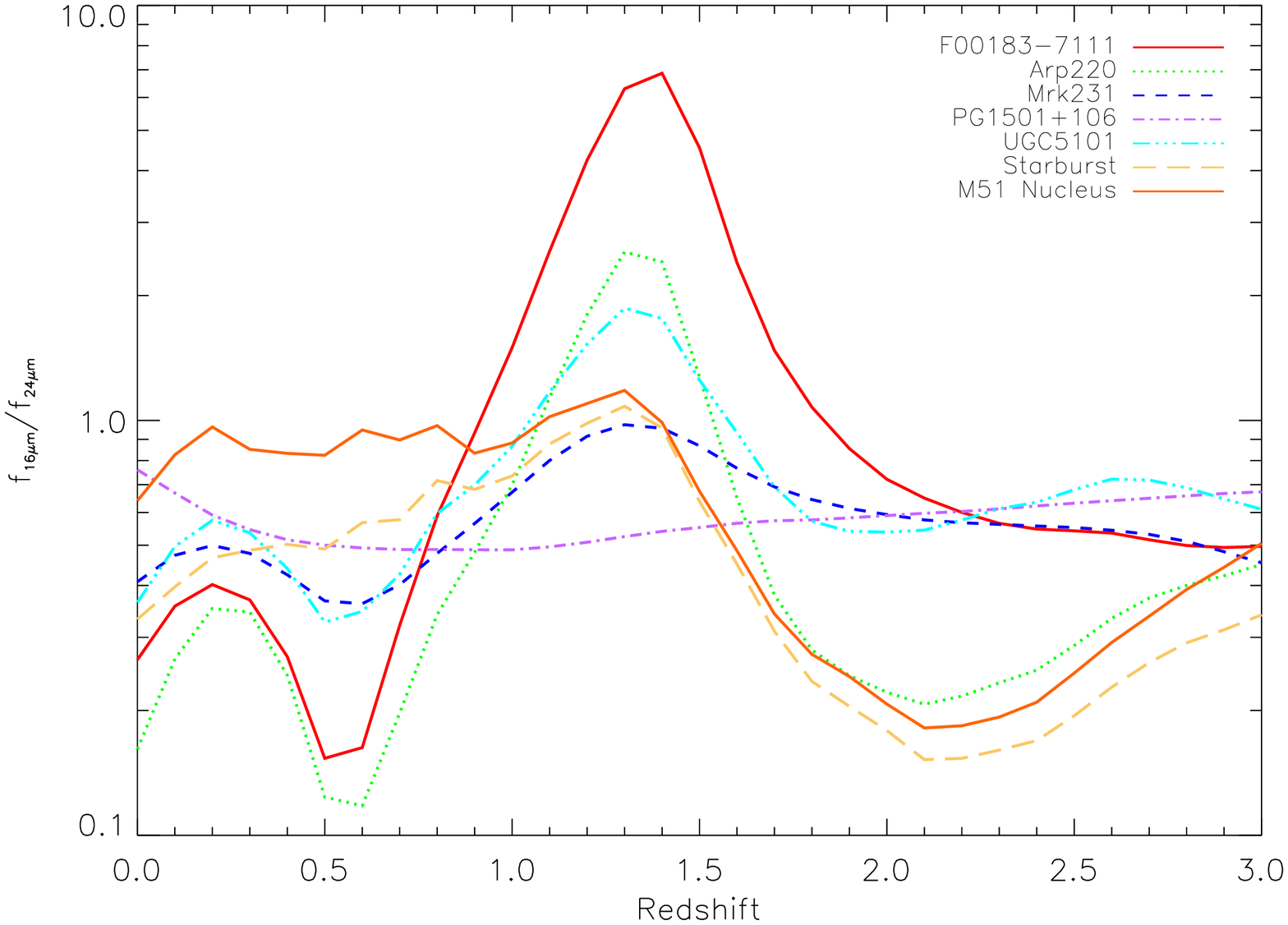}}
 \caption{The predicted ratio of the 16$\mu$m to 24$\mu$m flux
    densities as a function of redshift based on IRS spectra of
    template galaxies. We use the extreme silicate-absorption galaxy
    F00183-7111 \citep{Spoon04}, UGC5101 a ULIRG with considerable
    9.7$\mu$m absorption \citep{Armus04}, the prototypical AGN Mrk231
    \citep{Weedman05} and archetypal ULIRG Arp220 \citep{Armus05}, the
    typical quasar PG1501+106 from \citep{Hao05}, the average mid-IR
    SED of all starburst galaxies in the IRS GTO program from
    \citep{Brandl05}, as well as the nuclear spectrum of M51 available
    in the Spitzer archive.\label{ratios}}
\end{figure}

\clearpage

\begin{deluxetable}{cccccrccc}
\tabletypesize{\footnotesize}
\tablenum{1}
\footnotesize
\setlength{\tabcolsep}{0.1in}
\tablecaption{Catalog of the 16$\mu$m sources in the Bootes NOAO Wide Deep Field}
\tablewidth{0pc}
\tablehead{
\colhead{ID}  & \colhead{Spitzer Name\tablenotemark{a}}  & \colhead{f$_{16 \mu m}$}  & \colhead{$\sigma_{16 \mu m}$} & \colhead{f$_{24 \mu m}$\tablenotemark{b}} &  \colhead{$\frac{\rm f_{16 \mu m}}{\rm f_{24 \mu m}}$}  & \colhead{$B_W$}  &  \colhead{$R$}  &  \colhead{$I$}  \\
  &   & \colhead{(mJy)}  & \colhead{(mJy)} & \colhead{(mJy)}  &  & \colhead{(mag)}  &  \colhead{(mag)}  &  \colhead{(mag)}  }
\startdata
  1    &   SST16 J143413.49+332217.4    &      0.275    &      0.057    &    \nodata    &   $>$ 1.52    &   \nodata    &    24.72     &    23.65    \\
  2    &   SST16 J143408.45+332218.1    &      0.331    &      0.048    &     0.346     &       0.95    &    22.18     &    20.85     &    20.32    \\
  3    &   SST16 J143411.11+332212.2    &      0.237    &      0.048    &     0.207     &       1.14    &    24.82     &    23.74     &    23.00    \\
  4    &   SST16 J143411.37+332248.6    &      0.257    &      0.054    &     0.296     &       0.86    &    24.76     &    23.60     &    22.61    \\
  5    &   SST16 J143410.56+332302.7    &      0.212    &      0.050    &    \nodata    &   $>$ 1.18    &   \nodata    &   \nodata    &    24.81    \\
  6    &   SST16 J143411.18+332217.0    &      0.180    &      0.048    &    \nodata    &   $>$ 1.00    &   \nodata    &   \nodata    &   \nodata   \\
  7    &   SST16 J142820.50+353042.8    &      0.203    &      0.052    &     0.208     &       0.97    &   \nodata    &   \nodata    &    22.82    \\
  8    &   SST16 J142821.18+353137.1    &      0.166    &      0.054    &    \nodata    &   $>$ 0.92    &    25.49     &    23.32     &    22.00    \\
  9    &   SST16 J143212.04+351234.2    &      0.413    &      0.051    &     0.343     &       1.20    &    22.99     &    21.50     &    21.02    \\
 10    &   SST16 J143210.86+351339.4    &      0.251    &      0.053    &    \nodata    &   $>$ 1.39    &    23.36     &    22.44     &    22.07    \\
 11    &   SST16 J143030.36+343642.2    &      0.777    &      0.064    &     0.493     &       1.57    &    22.60     &    20.83     &    19.89    \\
 12    &   SST16 J143030.11+343704.3    &      0.220    &      0.061    &    \nodata    &   $>$ 1.22    &    25.14     &    23.70     &    22.94    \\
 13    &   SST16 J143028.50+343713.9    &      0.169    &      0.047    &     0.770     &       0.22    &    24.41     &    22.10     &    21.31    \\
 14    &   SST16 J143024.71+343645.6    &      0.242    &      0.054    &    \nodata    &   $>$ 1.34    &    26.79     &    23.76     &    22.70    \\
 15    &   SST16 J143029.44+343744.0    &      0.210    &      0.054    &     0.220     &       0.95    &    24.73     &   \nodata    &    23.26    \\
 16    &   SST16 J142743.41+342717.1    &      0.352    &      0.051    &    \nodata    &   $>$ 1.95    &    23.23     &    21.07     &    20.34    \\
 17    &   SST16 J142742.61+342639.8    &      0.181    &      0.049    &     0.569     &       0.31    &   \nodata    &    25.09     &    23.64    \\
 18    &   SST16 J142745.42+342654.2    &      0.142    &      0.044    &    \nodata    &   $>$ 0.79    &    22.88     &    21.45     &    20.88    \\
 19    &   SST16 J142748.11+342715.8    &      0.203    &      0.054    &     0.349     &       0.58    &    26.12     &    23.81     &    22.81    \\
 20    &   SST16 J142745.28+342644.5    &      0.174    &      0.048    &     0.277     &       0.62    &    25.35     &    25.15     &    24.66    \\
 21    &   SST16 J143109.78+343315.1    &      0.331    &      0.052    &    \nodata    &   $>$ 1.84    &    26.17     &    24.91     &    24.54    \\
 22    &   SST16 J143110.91+343325.0    &      0.210    &      0.051    &    \nodata    &   $>$ 1.16    &    24.44     &    22.39     &    21.64    \\
 23    &   SST16 J143107.75+343305.4    &      0.157    &      0.044    &    \nodata    &   $>$ 0.87    &    23.75     &    22.39     &    22.02    \\
 24    &   SST16 J142938.40+324348.0    &      0.955    &      0.074    &    \nodata    &   $>$ 5.31    &    22.95     &    20.56     &    19.78    \\
 25    &   SST16 J142936.66+324344.3    &      0.273    &      0.057    &     0.360     &       0.75    &    22.09     &    21.64     &    21.33    \\
 26    &   SST16 J142936.96+324420.2    &      0.426    &      0.069    &    \nodata    &   $>$ 2.37    &    25.06     &    23.98     &    23.28    \\
 27    &   SST16 J142936.64+324414.0    &      0.374    &      0.064    &    \nodata    &   $>$ 2.07    &    26.12     &    24.78     &    23.85    \\
 28    &   SST16 J143726.45+341935.8    &      0.150    &      0.047    &    \nodata    &   $>$ 0.83    &    23.95     &    22.12     &    21.36    \\
 29    &   SST16 J143726.18+341946.3    &      0.286    &      0.050    &     1.523     &       0.18    &    25.13     &    22.80     &    21.96    \\
 30    &   SST16 J143722.24+342005.5    &      0.336    &      0.057    &     0.561     &       0.59    &   \nodata    &   \nodata    &   \nodata   \\
 31    &   SST16 J143726.74+342034.7    &      0.594    &      0.053    &     0.562     &       1.05    &    22.75     &    20.93     &    20.25    \\
 32    &   SST16 J143006.09+341412.3    &      0.361    &      0.057    &    \nodata    &   $>$ 2.00    &    25.72     &    24.29     &    23.34    \\
 33    &   SST16 J143003.10+341447.2    &      0.260    &      0.051    &    \nodata    &   $>$ 1.44    &    18.55     &    16.67     &    17.74    \\
 34    &   SST16 J143005.05+341410.2    &      0.176    &      0.054    &     0.257     &       0.68    &    25.57     &    24.48     &    23.36    \\
 35    &   SST16 J143803.91+341458.3    &      0.655    &      0.059    &    \nodata    &   $>$ 3.63    &    25.00     &    22.58     &    21.38    \\
 36    &   SST16 J143808.17+341453.8    &      0.261    &      0.048    &    \nodata    &   $>$ 1.45    &    26.22     &    25.11     &    24.13    \\
 37    &   SST16 J143810.52+341500.1    &      0.186    &      0.056    &     0.225     &       0.82    &   \nodata    &   \nodata    &   \nodata   \\
 38    &   SST16 J143026.95+332006.1    &      0.228    &      0.052    &     0.473     &       0.48    &    26.68     &    25.55     &    24.30    \\
 39    &   SST16 J143546.89+343312.4    &      0.289    &      0.057    &     0.188     &       1.53    &    25.14     &    23.80     &    22.53    \\
 40    &   SST16 J143547.87+343320.5    &      0.241    &      0.055    &    \nodata    &   $>$ 1.34    &    24.52     &    23.23     &    22.51    \\
 41    &   SST16 J143543.08+343313.1    &      0.416    &      0.051    &    \nodata    &   $>$ 2.31    &    18.18     &    19.54     &    18.81    \\
 42    &   SST16 J143314.46+342452.9    &      0.901    &      0.065    &     0.301     &       2.99    &    22.99     &    21.03     &    20.04    \\
 43    &   SST16 J143309.84+342453.1    &      0.188    &      0.049    &     0.567     &       0.33    &    24.08     &    23.43     &    22.90    \\
 44    &   SST16 J143309.18+342445.2    &      0.181    &      0.053    &    \nodata    &   $>$ 1.00    &    21.92     &    20.65     &    20.04    \\
 45    &   SST16 J143312.72+342532.7    &      0.182    &      0.048    &     0.285     &       0.63    &    25.55     &    24.09     &    23.41    \\
 46    &   SST16 J142745.76+345418.0    &      0.304    &      0.060    &    \nodata    &   $>$ 1.69    &   \nodata    &    24.63     &    23.60    \\
 47    &   SST16 J143644.28+351107.9    &      0.226    &      0.063    &    \nodata    &   $>$ 1.25    &    23.83     &    21.47     &    20.84    \\
 48    &   SST16 J143642.85+351102.4    &      0.241    &      0.053    &     0.218     &       1.10    &    22.80     &    20.24     &    19.61    \\
 49    &   SST16 J143641.40+351144.9    &      0.309    &      0.065    &    \nodata    &   $>$ 1.71    &    25.09     &    23.41     &    22.54    \\
 50    &   SST16 J143640.08+351117.5    &      0.367    &      0.067    &    \nodata    &   $>$ 2.04    &    25.33     &    24.57     &    23.24    \\
 51    &   SST16 J143642.25+351057.5    &      0.290    &      0.052    &     1.149     &       0.25    &    24.31     &    22.14     &    21.41    \\
 52    &   SST16 J143027.91+343453.0    &      0.487    &      0.071    &    \nodata    &   $>$ 2.70    &    24.48     &    22.60     &    22.13    \\
 53    &   SST16 J143027.55+343453.9    &      0.546    &      0.075    &    \nodata    &   $>$ 3.03    &    24.25     &    22.10     &    21.13    \\
 54    &   SST16 J143809.03+342111.5    &      0.282    &      0.058    &     0.222     &       1.27    &    25.07     &    24.70     &    23.92    \\
 55    &   SST16 J143808.73+342043.0    &      0.211    &      0.058    &    \nodata    &   $>$ 1.17    &    25.42     &    24.45     &    23.46    \\
 56    &   SST16 J142649.81+333356.3    &      0.272    &      0.063    &     0.194     &       1.39    &    25.73     &    24.12     &    23.35    \\
 57    &   SST16 J142646.81+333400.4    &      0.325    &      0.058    &     0.314     &       1.03    &    24.61     &    23.51     &    22.85    \\
 58    &   SST16 J142649.10+333422.1    &      0.212    &      0.054    &     0.204     &       1.04    &    25.75     &    23.89     &    23.12    \\
 59    &   SST16 J142645.41+333442.7    &      0.293    &      0.071    &     0.236     &       1.24    &   \nodata    &    25.18     &    23.57    \\
 60    &   SST16 J142650.55+333454.8    &      0.245    &      0.057    &     0.274     &       0.89    &    22.62     &    20.91     &    20.46    \\
 61    &   SST16 J143510.36+335158.8    &      0.305    &      0.056    &     0.265     &       1.14    &    24.59     &    23.27     &    22.45    \\
 62    &   SST16 J143508.98+335237.7    &      0.237    &      0.055    &     0.253     &       0.93    &    23.79     &    21.69     &    21.01    \\
 63    &   SST16 J143508.52+335241.8    &      0.267    &      0.060    &     0.337     &       0.79    &    26.32     &    24.14     &    22.95    \\
 64    &   SST16 J143504.00+335215.0    &      0.796    &      0.063    &    \nodata    &   $>$ 4.42    &    22.63     &    20.53     &    19.87    \\
 65    &   SST16 J143506.65+335257.1    &      0.175    &      0.055    &    \nodata    &   $>$ 0.97    &    26.32     &    23.73     &    22.84    \\
 66    &   SST16 J142844.81+342936.7    &      0.423    &      0.074    &    \nodata    &   $>$ 2.35    &    23.40     &    22.30     &    21.27    \\
 67    &   SST16 J142828.48+354545.3    &      0.321    &      0.079    &     0.211     &       1.52    &    25.90     &    23.89     &    21.94    \\
 68    &   SST16 J142826.47+354635.9    &      0.271    &      0.073    &    \nodata    &   $>$ 1.50    &   \nodata    &   \nodata    &   \nodata   \\
 69    &   SST16 J142825.87+354636.1    &      0.249    &      0.071    &     0.215     &       1.15    &    24.67     &    21.81     &    20.33    \\
 70    &   SST16 J142826.03+354548.3    &      0.206    &      0.066    &     0.454     &       0.45    &   \nodata    &   \nodata    &   \nodata   \\
 71    &   SST16 J142826.12+354636.0    &      0.324    &      0.063    &     0.304     &       1.06    &   \nodata    &   \nodata    &    21.82    \\
 72    &   SST16 J142824.17+354709.6    &      0.305    &      0.076    &    \nodata    &   $>$ 1.69    &   \nodata    &   \nodata    &   \nodata   \\
 73    &   SST16 J142823.35+354624.8    &      0.305    &      0.077    &     0.415     &       0.73    &    26.83     &    24.31     &    22.86    \\
 74    &   SST16 J142607.37+351733.5    &      1.208    &      0.090    &    \nodata    &   $>$ 6.71    &    22.63     &    20.67     &    20.00    \\
 75    &   SST16 J142607.78+351653.7    &      0.602    &      0.079    &     0.270     &       2.22    &    23.18     &    20.84     &    20.13    \\
 76    &   SST16 J142608.56+351757.6    &      0.502    &      0.071    &     0.224     &       2.23    &    24.87     &    23.78     &    22.89    \\
 77    &   SST16 J143133.85+325922.6    &      0.369    &      0.068    &     0.457     &       0.80    &    25.47     &    23.51     &    22.38    \\
 78    &   SST16 J143135.30+330015.4    &      0.289    &      0.067    &     0.550     &       0.52    &    25.09     &    24.20     &    23.21    \\
 79    &   SST16 J143103.65+325619.5    &      0.501    &      0.080    &     0.183     &       2.73    &    22.77     &    20.95     &    20.02    \\
 80    &   SST16 J143100.54+325647.9    &      0.454    &      0.066    &    \nodata    &   $>$ 2.52    &    24.51     &    22.55     &    22.00    \\
 81    &   SST16 J142645.17+325814.0    &      0.201    &      0.058    &     0.307     &       0.65    &    22.33     &    20.79     &    20.18    \\
 82    &   SST16 J142646.22+325710.4    &      0.374    &      0.066    &    \nodata    &   $>$ 2.08    &    24.40     &    23.85     &    23.05    \\
 83    &   SST16 J142643.98+325702.9    &      0.332    &      0.065    &    \nodata    &   $>$ 1.84    &    25.62     &    23.99     &    23.02    \\
 84    &   SST16 J143516.07+330213.8    &      0.344    &      0.067    &    \nodata    &   $>$ 1.91    &    23.24     &    21.22     &    20.50    \\
 85    &   SST16 J143517.61+330208.4    &      0.468    &      0.069    &    \nodata    &   $>$ 2.60    &    24.19     &    22.37     &    21.22    \\
 86    &   SST16 J143517.91+330240.6    &      0.312    &      0.059    &     0.260     &       1.19    &    25.80     &    24.63     &    23.64    \\
 87    &   SST16 J143520.32+330206.2    &      0.174    &      0.052    &    \nodata    &   $>$ 0.97    &   \nodata    &   \nodata    &    24.76    \\
 88    &   SST16 J142632.88+332632.1    &      0.183    &      0.058    &     0.260     &       0.70    &    24.61     &    21.75     &    20.78    \\
 89    &   SST16 J142634.49+332640.4    &      0.227    &      0.056    &     0.190     &       1.19    &    27.00     &    26.22     &    25.19    \\
 90    &   SST16 J142636.19+332610.4    &      0.451    &      0.062    &    \nodata    &   $>$ 2.50    &    22.83     &    20.84     &    20.08    \\
 91    &   SST16 J143246.59+333038.1    &      0.245    &      0.064    &    \nodata    &   $>$ 1.36    &    24.31     &    22.61     &    21.56    \\
 92    &   SST16 J143248.82+333109.6    &      0.313    &      0.058    &     0.995     &       0.31    &    24.11     &    22.11     &    21.20    \\
 93    &   SST16 J143246.96+333019.8    &      0.581    &      0.066    &    \nodata    &   $>$ 3.23    &    23.13     &    20.97     &    20.27    \\
 94    &   SST16 J143046.24+333837.8    &      0.387    &      0.069    &    \nodata    &   $>$ 2.15    &    24.03     &    21.93     &    21.01    \\
 95    &   SST16 J143050.44+333857.0    &      0.617    &      0.058    &     0.307     &       2.00    &    25.04     &    22.62     &    21.93    \\
 96    &   SST16 J142913.17+333914.1    &      0.269    &      0.060    &    \nodata    &   $>$ 1.49    &    25.91     &    22.82     &    21.19    \\
 97    &   SST16 J142916.19+333834.1    &      0.391    &      0.057    &     0.384     &       1.01    &   \nodata    &    21.38     &    20.62    \\
 98    &   SST16 J143422.00+334014.7    &      0.297    &      0.061    &     0.826     &       0.36    &   \nodata    &   \nodata    &   \nodata   \\
 99    &   SST16 J143421.95+334018.7    &      0.454    &      0.060    &     0.579     &       0.78    &    23.47     &    21.36     &    20.59    \\
100    &   SST16 J142816.38+334052.6    &      0.507    &      0.060    &    \nodata    &   $>$ 2.81    &    25.00     &    23.41     &    22.29    \\
101    &   SST16 J143517.47+335952.8    &      0.255    &      0.051    &     0.297     &       0.85    &    25.63     &    24.55     &    24.33    \\
102    &   SST16 J143517.25+335857.3    &      0.272    &      0.059    &     0.237     &       1.14    &    25.01     &    23.91     &    23.10    \\
103    &   SST16 J143518.95+335924.0    &      0.482    &      0.063    &     0.410     &       1.17    &    23.41     &    21.21     &    20.63    \\
104    &   SST16 J143519.27+335859.4    &      0.244    &      0.075    &     0.428     &       0.57    &    25.47     &    24.14     &    23.36    \\
105    &   SST16 J143230.93+341812.3    &      0.283    &      0.062    &     0.426     &       0.66    &   \nodata    &   \nodata    &    25.17    \\
106    &   SST16 J143231.67+341755.1    &      0.247    &      0.062    &    \nodata    &   $>$ 1.37    &    25.33     &    23.40     &    22.33    \\
107    &   SST16 J143234.27+341759.3    &      0.254    &      0.055    &    \nodata    &   $>$ 1.41    &    24.12     &    22.60     &    21.90    \\
108    &   SST16 J142936.62+343547.2    &      0.255    &      0.054    &    \nodata    &   $>$ 1.42    &   \nodata    &   \nodata    &    23.81    \\
109    &   SST16 J142535.75+351336.6    &      0.203    &      0.061    &    \nodata    &   $>$ 1.12    &    26.08     &    23.79     &    22.34    \\
110    &   SST16 J142640.12+351436.2    &      0.362    &      0.063    &     0.612     &       0.59    &    24.10     &    23.35     &    22.64    \\
111    &   SST16 J142641.75+351356.8    &      0.397    &      0.063    &     0.297     &       1.33    &    25.33     &    23.50     &    22.26    \\
112    &   SST16 J142645.33+351416.0    &      0.398    &      0.059    &     0.380     &       1.04    &    23.83     &    21.78     &    20.88    \\
113    &   SST16 J142844.43+352716.3    &      0.423    &      0.061    &    \nodata    &   $>$ 2.35    &    23.82     &    22.48     &    21.40    \\
114    &   SST16 J142846.60+352701.9    &      0.411    &      0.057    &     0.502     &       0.81    &    26.77     &   \nodata    &    24.50    \\
115    &   SST16 J142846.31+352656.1    &      0.453    &      0.058    &     0.470     &       0.96    &    25.98     &    24.22     &    22.43    \\
116    &   SST16 J142849.45+352649.5    &      0.274    &      0.056    &     0.821     &       0.33    &    23.82     &    22.80     &    21.45    \\
117    &   SST16 J142844.80+352644.6    &      0.221    &      0.056    &     0.316     &       0.70    &    25.56     &    22.98     &    21.57    \\
118    &   SST16 J142921.96+321437.1    &      0.213    &      0.054    &     0.318     &       0.66    &   \nodata    &   \nodata    &   \nodata   \\
119    &   SST16 J142920.52+352837.2    &      0.235    &      0.056    &    \nodata    &   $>$ 1.30    &    25.51     &    24.35     &    23.22    \\
120    &   SST16 J143437.51+325837.8    &      0.183    &      0.053    &    \nodata    &   $>$ 1.01    &    24.98     &    22.51     &    21.33    \\
121    &   SST16 J143438.66+325743.3    &      0.288    &      0.056    &     0.418     &       0.69    &    22.62     &    21.00     &    20.47    \\
122    &   SST16 J143435.53+325739.7    &      0.382    &      0.057    &    \nodata    &   $>$ 2.12    &    24.71     &    23.86     &    22.64    \\
123    &   SST16 J142934.53+353055.4    &      0.392    &      0.056    &    \nodata    &   $>$ 2.18    &    26.08     &    24.27     &    22.51    \\
124    &   SST16 J143306.58+331721.6    &      0.182    &      0.056    &     0.666     &       0.27    &    22.09     &    21.14     &    20.81    \\
125    &   SST16 J143310.70+331704.5    &      0.321    &      0.054    &     0.204     &       1.56    &    23.14     &    21.93     &    21.42    \\
126    &   SST16 J143311.95+331649.9    &      0.392    &      0.066    &    \nodata    &   $>$ 2.18    &    22.43     &    20.97     &    20.42    \\
127    &   SST16 J142756.31+331646.1    &      0.837    &      0.069    &    \nodata    &   $>$ 4.65    &    22.95     &    21.10     &    20.30    \\
128    &   SST16 J143454.29+354403.5    &      0.529    &      0.063    &     0.314     &       1.68    &   \nodata    &   \nodata    &   \nodata   \\
129    &   SST16 J143456.49+354320.6    &      0.209    &      0.047    &    \nodata    &   $>$ 1.16    &    25.07     &    23.66     &    22.45    \\
130    &   SST16 J143348.90+332213.5    &      0.272    &      0.052    &    \nodata    &   $>$ 1.51    &   \nodata    &   \nodata    &    25.17    \\
131    &   SST16 J143350.47+332111.3    &      0.423    &      0.060    &    \nodata    &   $>$ 2.35    &    22.86     &    20.65     &    19.92    \\
132    &   SST16 J143346.77+332106.9    &      0.207    &      0.058    &    \nodata    &   $>$ 1.15    &    23.46     &    22.73     &    21.94    \\
133    &   SST16 J142951.88+322127.5    &      0.189    &      0.062    &     0.249     &       0.76    &    26.53     &    25.54     &   \nodata   \\
134    &   SST16 J142956.05+322126.5    &      0.227    &      0.058    &    \nodata    &   $>$ 1.26    &    22.47     &    21.23     &    20.80    \\
135    &   SST16 J142955.11+322046.1    &      0.383    &      0.060    &     0.338     &       1.13    &    24.41     &    21.80     &    21.06    \\
136    &   SST16 J143256.00+332947.0    &      0.273    &      0.063    &    \nodata    &   $>$ 1.51    &    24.41     &    23.77     &    23.02    \\
137    &   SST16 J143300.25+332945.8    &      0.196    &      0.053    &     0.340     &       0.57    &    24.80     &    23.73     &    22.79    \\
138    &   SST16 J143301.79+332927.5    &      0.360    &      0.065    &     0.190     &       1.89    &    24.38     &    23.48     &    22.62    \\
139    &   SST16 J143257.77+332943.6    &      0.221    &      0.057    &    \nodata    &   $>$ 1.23    &    26.28     &    24.62     &    23.55    \\
140    &   SST16 J143533.83+333718.9    &      0.296    &      0.059    &     0.520     &       0.56    &   \nodata    &    25.35     &   \nodata   \\
141    &   SST16 J143533.35+333656.1    &      0.249    &      0.064    &     0.753     &       0.33    &    26.11     &    24.46     &    23.56    \\
142    &   SST16 J143533.69+333632.4    &      0.296    &      0.059    &     0.300     &       0.98    &   \nodata    &    25.13     &    24.50    \\
143    &   SST16 J142950.93+334128.3    &      0.325    &      0.063    &     0.382     &       0.85    &    22.85     &    20.91     &    20.23    \\
144    &   SST16 J142952.67+334124.2    &      0.245    &      0.064    &    \nodata    &   $>$ 1.36    &    23.06     &    21.18     &    20.48    \\
145    &   SST16 J142953.79+334105.6    &      0.627    &      0.063    &     0.366     &       1.71    &    23.86     &    21.03     &    20.02    \\
146    &   SST16 J143245.10+334420.4    &      0.461    &      0.061    &     0.187     &       2.46    &    23.35     &    21.27     &    20.52    \\
147    &   SST16 J143241.61+334411.6    &      0.308    &      0.059    &    \nodata    &   $>$ 1.71    &    26.51     &    24.30     &    22.88    \\
148    &   SST16 J143641.92+350102.1    &      0.871    &      0.079    &    \nodata    &   $>$ 4.84    &   \nodata    &   \nodata    &   \nodata   \\
149    &   SST16 J143642.30+350153.3    &      0.278    &      0.057    &    \nodata    &   $>$ 1.54    &   \nodata    &   \nodata    &   \nodata   \\
150    &   SST16 J143640.05+350203.5    &      0.224    &      0.054    &     0.213     &   $>$ 1.24    &   \nodata    &   \nodata    &   \nodata   \\
\enddata
\tablenotetext{a}{SST16 source name derives from discovery with theIRS PU 16$\mu$m images; coordinates listed are in J2000; 16$\mu$m positions with typical 3\,$\sigma$ uncertainty of $\pm$ 1.2\arcsec; sources with an optical counterpart will also appear in NDWFS catalogs with prefix NDWFS and the optical source position; optical magnitudes are Mag-Auto from NDWFS Data Release Three, available at http://www.noao.edu/noao/noaodeep/); sources with a MIPS 24 $\mu$m counterpart will also appear in MIPS catalog.}
\tablenotetext{b}{Ratio of the 16$\mu$m and 24$\mu$m flux densities. Upper limits for 24$\mu$m sources are assumed to be 0.18 mJy.}
\end{deluxetable}

\end{document}